\documentclass[10pt,aps,showpacs,floatfix,amsmath,amssymb,groupedaddress]{revtex4}
\usepackage{amssymb}
\usepackage[dvips]{graphicx}
\usepackage[english]{babel}
\usepackage{indentfirst}
\usepackage{amsxtra}
\usepackage{amsmath}
\usepackage{supertabular}
\usepackage{multirow}
\usepackage[dvips]{color}
\usepackage [mathcal]{eucal}
\def\beq{\begin{equation}}
\def\eeq{\end{equation}}
\def\beqn{ \begin{eqnarray} }
\def\eeqn{ \end{eqnarray} }
\def\r{\mbox{\boldmath $r$}}

\def\p{\mbox{\boldmath $p$}}
\def\q{\mbox{\boldmath $q$}}
\def\s1s2{{ \boldsymbol{\sigma}(1) \cdot \boldsymbol{\sigma}(2) }}
\def\t1t2{{ \boldsymbol{\tau}(1) \cdot \boldsymbol{\tau}(2)  }}

\newcommand{\eep} { $(e,e^{\,\prime}p)$ }

\begin{document}
\noindent
\title{ Quasifree \eep reactions on 
        nuclei with neutron excess 
}

\author{C. Giusti, A. Meucci, F. D. Pacati}
\affiliation{
 \mbox { Dipartimento di Fisica Nucleare e Teorica, 
Universit\`a di Pavia, Pavia, ITALY} \\
\mbox {INFN, Sezione di Pavia,  Via Bassi 6, I-27100 Pavia, ITALY } \\
}
\author{G. Co', V. De Donno}
\affiliation{
 \mbox { Dipartimento di Fisica, Universit\`a del Salento, Lecce, ITALY} \\
\mbox {INFN, Sezione di Lecce, Via Arnesano, I-73100 Lecce, ITALY } \\
}
\date{\today}

\bigskip

\begin{abstract} 
We study the evolution of the \eep cross section on nuclei with 
increasing asymmetry between the number of neutrons and protons. 
The calculations are done within the framework of the distorted-wave 
impulse approximation, by adopting nonrelativistic and relativistic models.  
We compare the results obtained with three different approaches based on 
the mean-field description for the proton bound state wave function. 
In the nonrelativistic model phenomenological Woods-Saxon and Hartree-Fock 
wave functions are used, in the relativistic model the wave functions are 
solutions of Dirac-Hartree equations.
The models are first tested against experimental data on $^{16}$O, 
$^{40}$Ca, and $^{48}$Ca nuclei, and then they are applied 
to calculate \eep cross sections for a set of spherical calcium and oxygen 
isotopes.
From the comparison of the results obtained for the various 
isotopes we can infer information about the dependence of the various 
ingredients of the models on the neutron to proton asymmetry.  
\end{abstract}

\bigskip
\bigskip
\bigskip


\maketitle

\section{Introduction}
\label{sec:intro}

The understanding of the evolution of the nuclear properties with respect to the 
asymmetry between the number of neutrons and protons is one of the major topics 
of interest in modern nuclear physics.  It is going to extend our knowledge 
about the effects of isospin asymmetry on the nuclear structure, and is also 
relevant for the study of the origin and the limits of stability of matter in 
the universe. 

Nuclear reactions represent our main source of information on the
properties and on the structure of atomic nuclei.  Direct nuclear
reactions, where the external probe interacts with only one, or a few,
nucleons of the target nucleus, can give deep insight on the
single-particle (s.p.) properties of a many-body system.  In
particular, the \eep\ reaction, where a proton is emitted
with a direct knockout mechanism, represents a very clean probe to
explore thestructure pf the  proton-hole states of the nucleus
\cite{fru84,bof93,bof96}.

Electron scattering is probably the best tool for investigating the
structure of atomic nuclei and their constituents.  The
electromagnetic interaction is weak if compared with the strength of
the interaction between hadrons, therefore the nuclear many-body
system is only slightly perturbed by the probe. In addition, the
possibility of varying  independently the energy and momentum transferred
to the nucleus allows us to map the nuclear response as a function of
the excitation energy with a space resolution that can be adjusted to
the scale of processes we want to study. 
The theoretical description
of the electron-nucleus interaction is well controlled by the
perturbative quantum electrodynamics theory, and, for the
energies used in nuclear physics investigations, the one-photon
exchange approximation is usually enough to obtain a good description
of the electron scattering process. The advantages of electron
scattering have been exploited in the past to study the properties of
stable nuclei. These studies can now be extended to exotic nuclear
matter.

In the last thirty years a large amount of \eep\ data has provided
accurate informations on the
s.p. structure of stable closed-shell nuclei.  Many high-resolution
exclusive \eep\ experiments on several stable nuclei were carried out
at Saclay~\cite{fru84,bof93,bof96,mou76,ber82},
NIKHEF~\cite{dew90,lap93,bof93,bof96} and MAMI~\cite{blo95}.  Specific
quantum numbers and spectroscopic factors have been assigned to the
peaks in the energy spectrum by studying the missing energy and
momentum dependence of the experimental cross sections.  The data
analysis was carried out by using the program {\tt
DWEEPY}~\cite{giu87,giu88} which describes the process within the
theoretical framework of the nonrelativistic distorted-wave impulse
approximation (DWIA). In the program the effects of the final-state
interactions (FSI) due to the reinteraction of the emitted proton
with the remaining nucleus and also the distortion of the electron
wave functions produced by the nuclear Coulomb field are included.
For the data analysis, phenomenological ingredients were usually
adopted to calculate the s.p. bound and scattering states. In the original 
analyses of the experimental data, the outgoing nucleon scattering wave 
functions were eigenfunctions of an optical potential determined through a fit
to elastic unpolarized and polarized nucleon-nucleus scattering data.

The bound-state wave functions were calculated with a Woods-Saxon (WS)
well, where the radius was determined to fit the experimental momentum
distribution and the depth was adjusted to give the experimentally
observed separation energy of the bound final state. This theoretical
approach was able to describe, with a high degree of accuracy,  for a wide 
range of nuclei and for different
kinematics, the shape of the experimental momentum distributions at
missing-energy values corresponding to specific peaks in the energy
spectrum. In order to reproduce the size of the experimental cross
sections, the normalization of the bound-state wave function was
fitted to the data and identified with the spectroscopic factor. The
deviations of such normalization factors from the predictions of the
mean-field (MF) approximation are interpreted as
the effect of nucleon-nucleon correlations.

The description of \eep\ processes was done also with relativistic
DWIA (RDWIA) models
\cite{pick85,jin92,jin93,jin94,udi93,udi96,udi99,kel97,kel99,kel99a,hed95a,
joh96,meu01a,meu01,meu02a,meu02b,radici03,tamae09}.  In these
approaches the s.p. bound-state wave functions are obtained by solving
Dirac-Hartree equations with an interaction obtained by a relativistic
Lagrangian written in the context of the relativistic MF theory. The scattering
wave functions are obtained by solving the Dirac equation with
relativistic optical potentials determined by a fit to elastic
proton-nucleus scattering data.  Some of these models include also an
exact treatment of the Coulomb distortion of the electron waves
\cite{jin92,jin93,udi93,udi96,udi99}. The RDWIA calculations, which provide a 
good description of the old \eep\ data, are necessary for the analysis of
the more recent \eep\ data from JLab \cite{gao00,malov00}, measured in
kinematic conditions with higher 
values of the momentum transfer and of the outgoing proton energy,
unachievable in previous experiments.

In upcoming years the advent of radioactive ion beams facilities
\cite{tan95,gei95,mue01} will provide a large amount of data on
unstable nuclei. A new generation of electron colliders that use
storage rings is under construction at RIKEN (Japan)
\cite{sud01,kat03} and GSI (Germany) \cite{gsi06}. These facilities
will offer unprecedented opportunities to study the structure of
exotic unstable nuclei through electron scattering in the ELISe
experiment at FAIR in Germany \cite{elise} and the SCRIT project in
Japan \cite{sud10}.

Kinematically complete experiments, where all target-like reaction
products are detected, will become feasible for the first time,
allowing a clean separation of different reaction channels as well as
a reduction of the unavoidable radiative background seen in
conventional experiments. Therefore, even applications to stable
isotope will be of interest.

In this work we investigate how the models successfully used to
describe \eep\ data in stable nuclei behave when they are used to make
predictions on exotic nuclei. 
In our study, we consider both the nonrelativistic (DWIA) and 
the relativistic (RDWIA) approaches, and we apply
them to a set of calcium and oxygen isotopes.  We have chosen these
isotopes since data taken at NIKHEF for the doubly magic nuclei
$^{40}$Ca, $^{48}$Ca \cite{kra90t,kra01}, and $^{16}$O \cite{leu94} are
available. We first compare the performances of our models in
describing these data, then we apply them to some even-even isotopes
of these nuclei. We apply our models to  $^{40,48,52,60}$Ca  and 
$^{16,22,24,28}$O nuclei, where the s.p. levels
below the Fermi surface are fully occupied. In this manner we work
with spherical systems and minimize the pairing effects.  Our models
require the description of both ground and excited states of the
nuclear system, the latter one has a particle in the continuum. In
this article, we investigate how different descriptions, apparently
equivalent for stable nuclei, can produce different results when
applied to neutron rich nuclei.  
The DWIA calculations are carried out with the same code
{\tt DWEEPY} that was used for the analyses of the experimental data.
In the comparison with data we repeat the original analyses and
present the results of calculations performed under the same
conditions as in Refs. \cite{kra90t,kra01,leu94}. The results obtained
with phenomenological WS wave functions are compared with those
obtained by solving Hartree-Fock (HF) equations with Gogny-like
finite-range interactions.  The RDWIA calculations are performed with
the fully relativistic model developed in Ref. \cite{meu01a}.  The
theoretical investigation of exotic nuclei with models of proven
reliability in stable isotopes will test the ability of the
established nuclear theory in the domain of exotic nuclei, revealing 
the evolution of nuclear properties as a function of the asymmetry
between the number of neutrons and protons. Moreover, it will provide
valuable references for future experiments.

The paper is organized as follows. In Sec. \ref{sec:model}, we outline
the basic aspects of the models used for the calculations.  We present and discuss our results in
Sec. \ref{sec:results} and we draw our conclusions in
Sec. \ref{sec:conclusions}.


\section{Theoretical description of the \eep  process}
\label{sec:model}

Our description of the \eep\ reactions is based on the one-photon
exchange approximation, where the incident electron exchanges a
virtual photon, of momentum $\q$ and energy $\omega$, with the target
nucleus \cite{bof96}. In this approximation, the evaluation of the cross
section requires the calculation of the contraction between the lepton
tensor, that contains the electron current, and the hadron tensor,
that contains the nuclear electromagnetic
current. If we neglect the effects of the nuclear Coulomb field on the
electrons, the initial and final electron wave functions are plane
wave solutions of the Dirac equation. In this case, we can write the
lepton tensor as a kinematical factor, and the hadron tensor as a
bilinear product of the Fourier transforms of the transition matrix
elements of the nuclear electromagnetic current
operator between the initial $|\Psi_{\mathrm{i}}\rangle$ and final
$|\Psi_{\mathrm{f}}\rangle$ nuclear states
\begin{equation}
J^{\mu}({\mbox{\boldmath $q$}}) = \int \langle\Psi_{\mathrm{f}} 
|\hat{J}^{\mu}({\mbox{\boldmath $r$}})|\Psi_{\mathrm{i}}\rangle
\, 
{\mathrm{e}}^{\,{\mathrm{i}}{\footnotesize {\mbox{\boldmath $q$}}}\cdot 
{\footnotesize {\mbox{\boldmath $r$}}}}
d^3 r \,\,\,.     
\label{eq:jm}
\end{equation} 

If Coulomb-distorted electron wave functions are considered in the model,
the transition matrix element of the electron current replaces
the exponential in Eq.~(\ref{eq:jm}) and the calculation becomes much more
complicated \cite{bof96,giu87,giu88,udi93}. For the nuclei and the electron
energies considered in this work, however, a simple and accurate enough method 
to include the effects of Coulomb distortion is to preserve
the expression~(\ref{eq:jm}) but using an effective momentum transfer and 
change the momentum transfer into an
effective momentum transfer \cite{bof96,giu87,giu88,udi93}.  Therefore
in all the calculations presented in this paper we have used this approximation. 

The evaluation of the expression~(\ref{eq:jm}) requires a model that
describes the initial and final nuclear many-body states.  A detailed
description of our DWIA model, and of the related assumptions, can be found in 
Refs. \cite{bof82,bof96}. In short, we assume that in the final
state the residual nucleus, that is composed by $A-1$ nucleons, is left
in a bound state $| \Psi^{\mathrm B}_\alpha(E) \rangle$, characterized by the 
energy $E$ and the quantum numbers $\alpha$. Then, for the final nuclear 
state we select the channel subspace spanned by  the wave function 
$| \Psi^{\mathrm B}_\alpha(E) \rangle$, describing the $A-1$ nucleons, and a
function describing the state of the emitted nucleon. Moreover, we assume the
direct knockout mechanism, where the one-body nuclear electromagnetic current 
operator acts exclusively on the space spanned by this type of
many-body states.  

Under these assumptions we can write 
the expression~(\ref{eq:jm})  as
\begin{eqnarray}
\nonumber 
J^{\mu}({\mbox{\boldmath $q$}}) &=&
\int  
\langle \Psi_{\mathrm f}  |a^\dagger({\mbox{\boldmath $r$}})|
\Psi^{\mathrm B}_\alpha(E) \rangle \,
\hat{\jmath}^{\mu}({\mbox{\boldmath $r$}}) \, 
\langle\Psi^{\mathrm B}_\alpha(E)|a({\mbox{\boldmath $r$}})
|\Psi_{\mathrm i}\rangle \,
{\mathrm{e}}^{\,{\mathrm{i}} {\footnotesize {\mbox{\boldmath $q$}}}\cdot
{\footnotesize {\mbox{\boldmath $r$}}}} 
\, d^3r \\
&=&
\int  \chi^{(-)*}_{E\alpha} ({\mbox{\boldmath $r$}}) 
\, \hat{\jmath}^{\mu} ({\mbox{\boldmath $r$}}) \,\phi_{E\alpha}({\mbox{\boldmath $r$}})
\left[S_\alpha(E)\right]^{1/2} \, 
{\mathrm{e}}^{\,{\mathrm{i}} {\footnotesize {\mbox{\boldmath $q$}}}\cdot
{\footnotesize {\mbox{\boldmath $r$}}}} 
\, d^3r \,\,\,,
\label{eq:dwia}
\end{eqnarray}
where we have not explicitly indicated the dependence on the spin and
isospin variables and we understand
that the one-body electromagnetic current operator
$\hat{\jmath}^{\mu}$ acts on the nuclear wave fuctions only.  In
Eq.~(\ref{eq:dwia}) we have indicated with $a(\r)$ the operator which
annihilates a nucleon with coordinate $\r$, and with
$\chi^{(-)}_{E\alpha}$ the distorted wave function of the emitted
nucleon
\begin{equation}
\langle\Psi^{\mathrm B}_\alpha(E)|a({\mbox{\boldmath $r$}})|
\Psi_{\mathrm f}\rangle
= \chi^{(-)}_{E\alpha}({\mbox{\boldmath $r$}}) 
\,\,\,.
\label{eq:dw}
\end{equation}
In the initial state for the wave function of the bound nucleon we have defined 
the overlap function, which describes the residual nucleus as a hole state in the 
target, as
\begin{equation}
\langle\Psi^{\mathrm B}_\alpha(E)|a({\mbox{\boldmath $r$}})
|\Psi_{\mathrm i}\rangle
=
\left[S_\alpha(E)\right]^{1/2}\phi_{E\alpha}({\mbox{\boldmath $r$}}) 
\,\,\,.
\label{eq:ovf}
\end{equation}
The overlap function contains the effects of nuclear 
correlations \cite{bof96,ant88,ant93,dic04}.
In the definition Eq.~(\ref{eq:ovf}) the function $\phi_{E\alpha}(\r)$ is 
normalized to unity. The spectroscopic factor $S_\alpha(E)$ is the norm of the 
overlap function and gives the probability of removing from the target a 
nucleon at $\r$ leaving the residual nucleus in the state 
$\Psi^{\mathrm  B}_\alpha(E)$.

This model can be formulated in both the 
DWIA and the RDWIA. In RDWIA calculations four-vector relativistic wave
functions for the initial bound and the final scattering states and,
coherently, relativistic expressions for the nuclear current operator
are used \cite{meu01a}.  We point out that in our DWIA calculations
some relativistic corrections are included in the kinematics and in
the nuclear current operator.

Formally, as described in Refs. \cite{bof82,bof96},  the s.p. scattering and 
bound state wave functions are derived in the model of
Eq.~(\ref{eq:dwia}) as eigenfunctions of an energy-dependent non-Hermitean 
Feshbach-type optical-model
Hamiltonian. Such a consistent treatment is extremely
difficult and, at present, bound and scattering states calculated from 
the theoretical optical potential are not available. For this reason, in 
actual calculations phenomenological wave functions are usually employed.
In the nonrelativistic DWIA calculations, the outgoing proton wave
functions are eigenfunctions of the complex phenomenological
energy-dependent and A-dependent optical potential of Ref.
\cite{sch82}, that contains a central and a spin-orbit term, and also
a term dependent on the 
nuclear asymmetry $(N-Z)/A$. This is the same
optical potential used in the original analyses of the NIKHEF data
\cite{leu94,kra90t,kra01}. We used this potential in all the DWIA
calculations done for the various isotopes we have investigated.

In our RDWIA model, the ejectile wave function is written in terms of
its positive energy component $\Psi_{f+}$ following the direct Pauli 
reduction method~\cite{hed95}. This scheme appears
simpler and it is equivalent to the solution of the Dirac equation. 
The resulting Schr\"odinger-like equation for $\Psi_{f+}$
contains equivalent nonrelativistic central and spin-orbit
potentials which are functions of the relativistic, energy-dependent,
scalar and vector potentials. The Darwin nonlocality factor, that
contains the effect of the negative-energy components of
the spinor, is reabsorbed in the current operator, which becomes
an effective relativistic one-body operator depending
on the Dirac scalar and vector potentials as well as on the prescription for 
the electromagnetic current.

  In our calculations we used the
relativistic EDAD1 potential of Ref. \cite{coo93}, constructed to fit
proton elastic scattering data on several nuclei in an energy range up
to 1040 MeV.

The s.p. bound-state wave functions
$\phi_{E\alpha}$ have been obtained in a MF model, and the values of
the spectroscopic factors have been set to unity. In this article we
intend to study the sensitivity of our results to the changes of the
hole s.p. wave function. To this aim, we have performed calculations
for the various nuclei under investigation with different MF
approaches.

In a first approach, for the DWIA calculations, we repeat for the
comparison with data the original analyses of
Refs. \cite{kra90t,kra01,leu94} and adopt the same phenomenological
ingredients. As indicated in the introduction, in these calculations
the hole wave functions are calculated by using a WS well where the
values of the depth and of the radius are adjusted to reproduce,
respectively, the experimentally observed separation energy of the
bound final state and the width of the experimental momentum
distributions. The values of the parameters used in the calculations
can be found in Refs. \cite{kra90t,kra01,leu94}. In order to reproduce
the magnitude of the experimental cross sections a multiplicative
reduction factor is applied to the calculated cross sections. These
factors, identified with the spectroscopic factors, indicate that in
\eep\ reactions the removal of the MF s.p.
strength for quasi-hole states
near the Fermi energy is about 60-70\% of the prediction of the
s.p. model \cite{bof96,lap93,kra01}.  
This information from \eep experiments is
up to now limited to stable isotopes.

The source of the reduction of the \eep\ spectroscopic factor with
respect to the MF value has been investigated by using various
methodologies which consider different types of correlations, i. e.,
effects beyond the MF model.  The short-range and tensor correlations,
which arise from the characteristics of the bare nucleon-nucleon
interaction, account for a reduction factor of at most 10-15\%
\cite{gai00,iva01,mue95,van98,roh04,sub08}. The remaining, and
larger, part of the quenching is due to long-range correlations
related to the coupling between s.p. motion and
collective surface vibrations \cite{dic04,bar09}.

In a second approach, also used in DWIA calculations, the hole
s.p. wave functions are obtained by solving HF equations with the
technique presented in Refs. \cite{co98b,bau99}.  In these
calculations we use two different parameterizations of the
finite-range Gogny interactions, the more traditional D1S force
\cite{ber91} and the new D1M force \cite{gor09}. The differences
between the results obtained with these two forces are rather small
when compared with the differences 
with the results obtained with the other methods. For
this reason we shall present here only the results obtained with the
D1M interaction, which produces a neutron matter equation of state
which has a plausible behavior at high densities, in contrast to that
obtained with the D1S interaction.

In the third approach we made RDWIA calculations where the hole wave
functions are obtained in the context of the relativistic MF
approach by solving
the Dirac-Hartree equations.  The nucleon interaction is derived from
a relativistic Lagrangian containing $\sigma$, $\omega$ and $\rho$
meson fields and also the photon field.  The nuclear and Coulomb
potentials are obtained by solving self-consistently the Klein-Gordon
and Maxwell equations. This approach satisfactorily reproduces global
and s.p. properties of several nuclei \cite{hor79,hor81,hor91}.

The last ingredient necessary to calculate the
expression~(\ref{eq:dwia}) is the current $\hat{\jmath}^{\mu}$.  In RDWIA
calculations the electromagnetic one-body current operator corresponds
to the relativistic current-conserving $cc2$ expression of
Ref. \cite{def83}. This expression is consistent with the current we
use in DWIA calculations, where the relativistic corrections up to
order $1/M^2$, where $M$ is the nucleon mass, are obtained from a
Foldy-Wouthuysen transformation applied to the interaction Hamiltonian
where the nuclear current has the same form as in the $cc2$
expression.

The values of some quantities related to the wave functions and density
distributions, obtained in the different MF approaches for the nuclei under
investigation, are compared in Tab.~\ref{tab:oisot} for the oxygen isotopes 
and in Tab.~\ref{tab:caisot} for the calcium isotopes. 
In these tables the proton separation energies and  the root mean squared (rms) radii 
of the bound state wave functions 
are shown. The proton separation energies of the WS calculations reproduce the
experimental values. 
For $^{28}$O and $^{60}$Ca nuclei, where no experimental values are available, 
the depth of the WS well has been chosen to reproduce the separation energy 
obtained in the nonrelativistic HF approach. We remark that, experimentally, 
$^{28}$O and $^{60}$Ca nuclei are unbound; however our MF calculations bind 
all the nuclei we have investigated.
 
In Tab.~\ref{tab:oisot} and \ref{tab:caisot} we also present the rms radii of 
the global proton and neutron density distributions for the HF and the 
relativistic models. In the phenomenological WS approach the parameters of the
WS well are determined to reproduce the experimental separation energy and 
the width of the experimental \eep distribution and can give information only 
on the wave function of the considered proton. 
The behavior of the neutron rms radii is rather obvious
and increases with the increasing number of neutrons. In the HF model the
neutron radii are slightly smaller than in the relativistic model, but for 
$^{16}$O and $^{40}$Ca, where they are practically the same.
The behavior of the proton rms radii is less obvious. The proton number is the
same in each isotope chain, but in the HF case we observe a small increase of
the proton rms radius with  the increasing number of neutrons. 
In contrast, in the relativistic case there is a slight decrease of the 
radius from $^{16}$O  to $^{22}$O and from $^{40}$Ca  to $^{48}$Ca, then for 
nuclei with larger neutron excess the radius increases. In general in the HF 
approach the proton radii are a bit larger than in the relativistic approach. 

The modifications of the proton radii due to the presence of neutrons
are related to the proton-neutron interaction, which is responsible also for the
value of the proton-neutron symmetry energy in nuclear matter.
In HF calculation we obtain a value of 29.45 MeV while that of the
relativistic calculation is 35 MeV. These values should be compared
with an empirical value of  32$\pm$3 MeV.

The experimental data of the \eep\ reaction are separated in different
peaks corresponding to specific values of the missing energy
$E_{\mathrm m} = \omega-T'-T_{\mathrm B}= S_{\mathrm p} + E_{\mathrm
x}$, where $T'$ and $T_{\mathrm B}$ are the kinetic energies of the
outgoing nucleon and of the residual nucleus, respectively,
$S_{\mathrm p}$ is the nucleon separation energy, and $E_{\mathrm x}$
is the excitation energy of the residual nucleus. For each peak, the
data are usually presented in terms of the reduced cross section as a
function of the missing momentum $p_{m}=|{\mbox{\boldmath $p$}}_{\mathrm m}|$, which is the magnitude of the recoil momentum of
the residual nucleus \cite{bof96,dew90,lap93}. The reduced cross
section is the cross section divided by a kinematical factor and by the
elementary off-shell electron-proton scattering cross section. For the
latter cross section, the $cc1$ prescription of Ref. \cite{def83} is
usually adopted. In the reduced cross section the complicated
dependence of the cross section on the kinematic variables is reduced
to a twofold function of $E_{\mathrm m}$ and $p_{\mathrm m}$.

If we neglect FSI the wave function of the emitted proton is a plane
wave. In this plane-wave impulse approximation (PWIA), the missing
momentum ${\mbox{\boldmath $p$}}_{\mathrm m}$ corresponds, apart for a
minus sign, to the initial momentum of the emitted nucleon in the
nucleus. In the PWIA the cross section is factorized into the product of a
kinematical factor, the elementary off-shell electron-proton
scattering cross section, and the hole spectral function.  Thus, in the 
PWIA the reduced cross section is the squared Fourier transform of the
hole wave function, and can be interpreted as the momentum
distribution of the emitted proton when it was inside the nucleus.
This factorization is destroyed in the DWIA by FSI. However, even in the DWIA,
the reduced cross section is an interesting quantity that can be
regarded as the nucleon momentum distribution modified by FSI.

The theoretical approaches outlined above have been used to calculate
\eep\ reduced cross sections in the parallel and perpendicular
kinematics selected in the experiments.  In the so-called parallel
kinematics \cite{bof96}, the momentum of the outgoing proton $\p'$ is
kept fixed and taken parallel, or antiparallel, to the direction of
the momentum transfer $\q$.  Different values of the missing momentum
$p_{\mathrm m}$ are obtained by varying the electron scattering angle
and, as a consequence, $q$.  In the so-called perpendicular, or
($\q,\omega$) constant kinematics, the momentum transfer $\q$ and the
outgoing proton momentum $\p'$ are kept constant and the value of the
missing momentum $p_{\mathrm m}$ is changed by varying the angle of
the outgoing proton.


\section{Results}
\label{sec:results}

In this section we present the results of our calculations of the
exclusive \eep\ cross sections for a set of oxygen and calcium
isotopes. We compare the results obtained by using the different
nuclear structure models described in the previous section.  First we
show the performances of the various models in the description of the
existing experimental data in $^{40}$Ca, $^{48}$Ca, and $^{16}$O
nuclei. Then we present the results obtained for the other isotopes.

\begin{figure}
\begin{center}
\includegraphics[height=12cm, width=12cm]{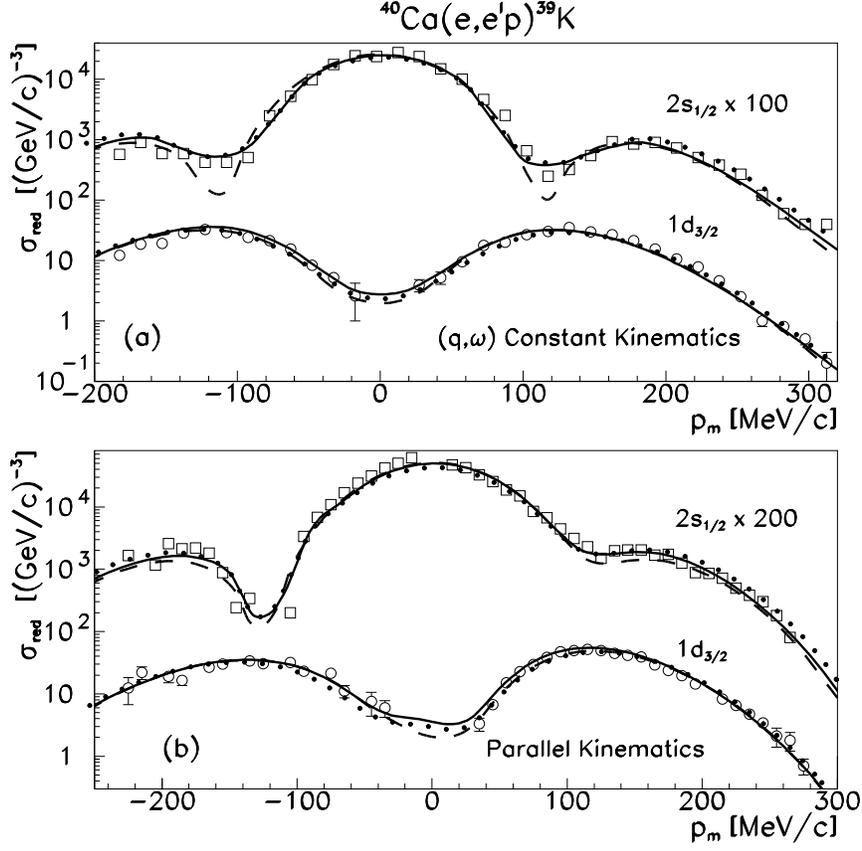} 
\caption{\small Reduced cross sections of the
  $^{40}$Ca\eep\ reaction as a function of the missing
  momentum $p_{\mathrm m}$ for the transitions to the $3/2^{+}$ ground
  state and to the $1/2^{+}$ excited state at 2.522 MeV of $^{39}$K.
  In panel (a) we show the results obtained in ($\q,\omega$) constant
  kinematics, with incident electron energy $E_0= 483.2$ MeV, electron
  scattering angle $\vartheta= 61.52^\circ$, and $q=450$ MeV/$c$. In panel (b) 
  we show the results obtained in parallel kinematics, with $E_0= 483.2$ MeV. 
  The outgoing proton energy is $T'= 100$ MeV in both kinematics.
  The experimental data are taken from Ref. \cite{kra90t}.  
  The solid lines give the DWIA-WS results, the dotted lines the DWIA-HF
  results, and the dashed lines the RDWIA results. The meaning of the
  negative values of $p_{\mathrm m}$ is explained in the text.
  The theoretical results have been multiplied by the reduction factors
  presented in Tab.~\ref{tab:redfac}.  
}
\label{fig:ca40exp}
\end{center}
\end{figure}

Measurements for the $^{40}$Ca\eep reaction were carried out at NIKHEF
in both parallel and ($\q,\omega$) constant kinematics \cite{kra90t}.
The comparison with the experimental reduced cross sections in both
kinematics is displayed in Fig.~\ref{fig:ca40exp}.  We have considered
the transitions to the $3/2^{+}$ ground state of the $^{39}$K nucleus,
corresponding to the knockout of the proton from the $1d_{3/2}$
s.p. level, and to the $1/2^{+}$ excited state of the $^{39}$K nucleus
at $E_{\mathrm x}= 2.522$ MeV, obtained by knocking out a proton from
the $2s_{1/2}$ s.p. level.  In Fig.~\ref{fig:ca40exp}, as well as in
the subsequent figures, positive (negative) values of $p_{\mathrm m}$
refer to situations where in ($\q,\omega$) constant kinematics the
angle between the outgoing proton momentum $\p'$ and the incident
electron $\p_0$ is larger (smaller) than the angle between $\q$ and
$\p_0$. In parallel kinematics positive and negative values of
$p_{\mathrm m}$ indicate, respectively, the $|\q|<|\p'|$ and
$|\q|>|\p'|$ cases. 

All the theoretical results shown in the figure provide a good
description of the experimental data.  As in the original data analysis
\cite{kra90t}, for the WS wave functions the radius of the potential
was chosen to reproduce the width of the experimental distribution.
On the other hand, no free parameters are used in the nonrelativistic
HF and in the relativistic Dirac-Hartree wave functions, that are able
to give an equivalently good description of data. 

In order to reproduce the magnitude of the experimental data, a
reduction factor has been applied in Fig.~\ref{fig:ca40exp} to all the
theoretical results. These factors, listed in Tab.~\ref{tab:redfac},
have been determined by a fit of the calculated reduced cross sections
to the data over the whole missing-momentum range considered in the
experiment.  The reduction factors applied to the DWIA-WS results are
identical to those obtained in the data analysis of
Ref. \cite{kra90t}, where the same optical potential and WS wave
functions were used.
 
In RDWIA calculations the $cc2$ expression for the current 
operator \cite{def83} has been used. Different expressions can give an 
equivalently good agreement with the shape of the experimental momentum 
distribution,  but somewhat different reduction factors must be applied to
reproduce the magnitude of the experimental reduced cross sections
\cite{udi93,radici03}. We checked that
the reduction factors are about 10\% smaller when the $cc1$ current is 
used, and about 10\% larger with the $cc3$ current.

In all the calculations the reduction factors obtained for the
transition to the $3/2^{+}$ ground state in perpendicular kinematics
are about 20-25\% lower than those obtained in parallel kinematics.
The source of this difference is not clear. It may have an
experimental motivation, or it may be due to effects that give
different contributions in different kinematics and that are not
adequately considered in the model. In any case, the difference
reflects the uncertainties in the identification of the spectroscopic
factor as a simple reduction factor of the theoretical results with
respect to the experimental data.
   
\begin{figure}
\begin{center}
\includegraphics[height=9cm, width=9cm]{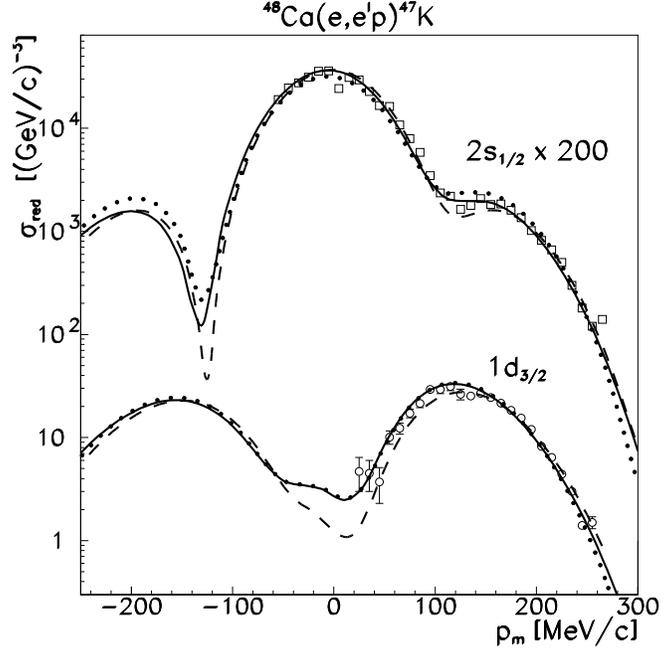} 
\caption{\small Reduced cross section of the $^{48}$Ca\eep\ reaction
as a function of $p_{\mathrm m}$ for the transitions to the $1/2^{+}$
ground state and to the $3/2^{+}$ excited state at 0.36 MeV of
$^{47}$K in parallel kinematics, with $E_0= 440$ MeV and $T'= 100$
MeV. The line convention is the same as in Fig.~\ref{fig:ca40exp}. The
experimental data are from Ref. \cite{kra90t}. The values of the
reduction factors multiplying the theoretical results are given in
Table \ref{tab:redfac}. 
}
\label{fig:ca48exp}
\end{center}
\end{figure}

Measurements for the $^{48}$Ca\eep\ reaction were carried out at NIKHEF in
parallel kinematics \cite{kra90t}. Thus, for calcium isotopes, we have the 
opportunity to test the evolution of our theoretical input against the change of  
neutron number in comparison with the experimental data. The comparison 
between our DWIA and RDWIA results and the $^{48}$Ca\eep\  data is shown in 
Fig.~\ref{fig:ca48exp} for the transitions to the $1/2^{+}$
ground state and to the first $3/2^{+}$ excited state at $E_{\mathrm
x}= 0.36$ MeV of $^{47}$K.

As in the case of $^{40}$Ca, all the theoretical results give a very good 
description of the shape of the experimental momentum distribution. 
Also in this case, reduction factors have been applied to all the calculated 
reduced cross sections presented in the figure. These factors are listed 
in Tab.~\ref{tab:redfac}.  
For the DWIA-WS results the reduction factors are identical to those obtained 
in the data analysis of Ref. \cite{kra01}.
We notice that for all our model calculations the reduction factors  obtained 
for the $1d_{3/2}$ state of $^{48}$Ca are consistently lower than those 
obtained for the $1d_{3/2}$ state of $^{40}$Ca in the same parallel kinematics.

\begin{figure}
\begin{center}
\includegraphics[height=9cm, width=9cm]{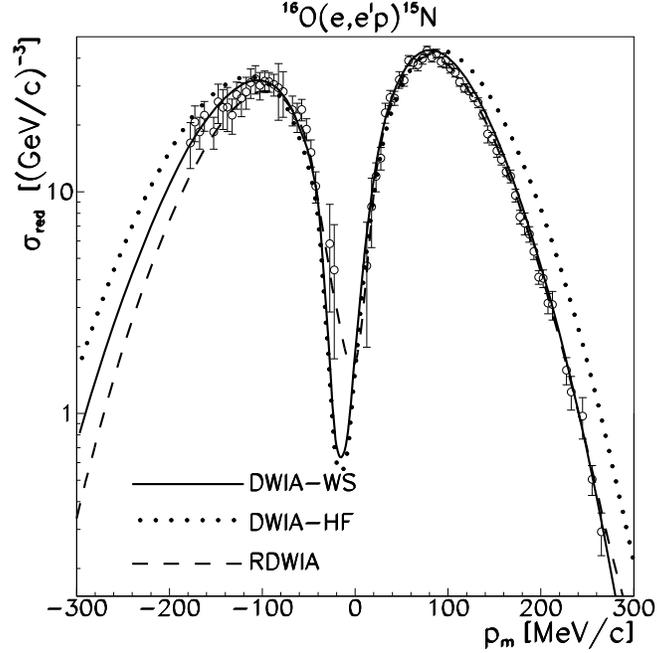} 
\caption{\small Reduced cross section of the $^{16}$O\eep\ reaction as
  a function of  $p_{\mathrm m}$ for the
  transition to the $1/2^{-}$ ground state of $^{15}$N in parallel
  kinematics with $E_0= 520.6$ MeV and $T'= 90$ MeV.  The
  experimental data are from Ref. \cite{leu94}. The values of the
reduction factors multiplying the theoretical results are given in
Table \ref{tab:redfac}.
}
\label{fig:oexp}
\end{center}
\end{figure}

In Fig.~\ref{fig:oexp} we compare our results with the
$^{16}$O\eep\ data measured at NIKHEF in parallel
kinematics \cite{leu94} for the transition to the $1/2^{-}$ ground
state of $^{15}$N, which is obtained by knocking out a proton from the 
 $1p_{1/2}$ s.p. level. Also in this case, the results of our calculations
are multiplied by the reduction factors given in Tab.~\ref{tab:redfac}. 
The DWIA-WS reduced cross section gives a good description of the data. As in 
the calcium cases, the value of the reduction factor is the same obtained in 
the data analysis of Ref. \cite{leu94}, where the same  bound-state wave 
function and optical potential were used.

Also the RDWIA calculation gives a good description of the data. This 
is the same result presented in \cite{meu01a}, and also the same
reduction factor has been applied. The situation is slightly different
for the DWIA result obtained with the HF wave function. In this
case, the description of the shape of the experimental momentum
distribution is slightly worse. We observe that the width of the HF
distribution is larger than that shown by the data and well reproduced
by the other calculations. This indicates that the rms radius of the
$1p_{1/2}$ HF wave function is smaller than those of the WS and
relativistic wave functions (see Tab.~\ref{tab:oisot}).  
We observe in Tab.~\ref{tab:redfac} that the reduction factor 0.89, that has 
been applied to the DWIA-HF result to reproduce the magnitude of the 
experimental data in the maximum region, is remarkably larger than those 
applied to the DWIA-WS and RDWIA results. We notice that this large 
value has been obtained fitting the maximum of the experimental reduced cross 
section, and not through a best fit to the whole momentum distribution like 
with DWIA-WS and RDWIA. Moreover, a poor description of the experimental 
shape does not permit a reliable determination of the spectroscopic factor. 

Our nuclear models have been used to calculate \eep\ cross sections
for other oxygen and calcium isotopes. We intend to investigate how
the characteristics of the s.p. proton wave functions, describing the
initial hole state, evolve as a function of the neutron number and how
this evolution affects the \eep\ cross sections.

Once the effective interaction is chosen, the HF and the relativistic
Dirac-Hartree approaches are parameter-free, and automatically provide
s.p. energies and wave functions for each nucleus. The situation is
different in the phenomenological approach, where the parameters of
the WS wells change for each nucleus.  We changed the radius of the
well by using the empirical $R\,=\,r_o\, A^{1/3}$ rule.  Moreover, the
depths of the wells are determined to obtain the experimental
separation energies taken from the compilation of
Refs. \cite{aud03,audw,nndc}.

We first discuss the results obtained for the calcium isotopes.  We
have considered the $^{40,48,52,60}$Ca isotopes, since they are
spherical nuclei where the s.p. levels are fully occupied and the
pairing effects are negligible.  For all these isotopes we have
considered proton knock-out emission from the $1d_{3/2}$ and
$2s_{1/2}$ states. Our results are shown in Fig.~\ref{fig:caiso_par}
for parallel kinematics and in Fig.~\ref{fig:caiso_per} for
($\q,\omega$) constant kinematics. The values of all the kinematic
variables are the same as in Figs.~\ref{fig:ca40exp} and
\ref{fig:ca48exp}. The only difference is that, since the separation
energy of the proton increases with increasing number of neutrons, we
changed, for the different isotopes, the energy of the outgoing
electron, and consequently the energy transfer, in order to keep
constant, via energy conservation, the energy of the outgoing
proton. We have verified that these differences in the kinematic
variables do not produce significant effects on the final results.

In  Figs.~\ref{fig:caiso_par} and \ref{fig:caiso_per} we
compare the reduced cross sections calculated for the various isotopes
obtained by using the three different nuclear models: DWIA-WS, DWIA-HF,
and RDWIA.  The comparison illustrates how these models describe
the effects produced by the asymmetry between the number of neutrons
and protons. No reduction factors have been applied to the curves
presented in these figures.  

It is interesting to remark that the evolution of the cross
section with respect to the change of the neutron number is
the same in all the panels of both figures. We observe that the
$^{40}$Ca lines are always above the other ones, and the size of
the curves decreases with the increasing number of neutrons. This
behavior is clearer in the DWIA-WS results, panels (a) and (b) of
both figures, and becomes less evident in the other cases,
especially in the RDWIA ones. 

\begin{figure}
\begin{center}
\includegraphics[height=15cm, width=15cm]{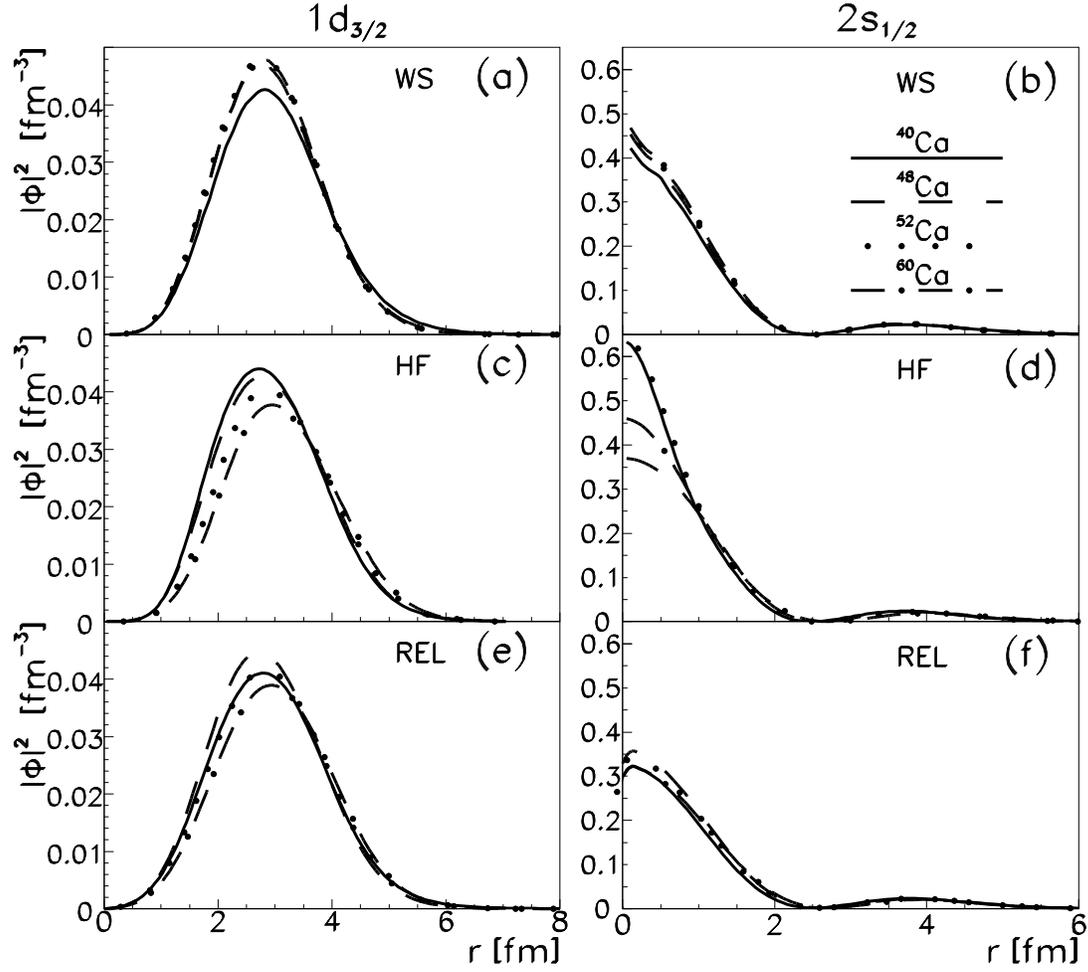} 
\caption{\small 
Squared moduli of the radial part of the $1d_{3/2}$ (left panels) and
$2s_{1/2}$ (right panels) s.p. wave functions for $^{40}$Ca (solid lines), 
$^{48}$Ca (dashed lines), $^{52}$Ca (dotted lines), and $^{60}$Ca (dot-dashed 
lines). In panels (a) and (b) we show the WS wave functions, in
panels (c) and (d) the HF wave functions, and in
panels (e) and (f) the relativistic wave functions obtained in the 
Dirac-Hartree approach.  The normalization of the curves is 
$\int \,dr\, r^2\, |\phi|^2 = 1$.  }
\label{fig:densca}
\end{center}
\end{figure}

For a better understanding of these results it is interesting to
consider the s.p. hole wave functions obtained in the three models. We
show in Fig.~\ref{fig:densca} the squared moduli of the radial part of
the s.p wave function for the $1d_{3/2}$ and $2s_{1/2}$ states of the
various calcium isotopes.  The relativistic wave functions shown in
the figure are obtained by summing the squared of the radial part of
the upper and lower components of the Dirac spinor. All the curves shown in
the figure are normalized to one.

The behavior of the WS wave functions shown in panels (a) and (b) can be 
understood by considering that the depth of the WS
well becomes deeper with increasing neutron number. This
deepening occurs because by increasing the neutron number the proton 
experiences more binding, its separation energy increases, and  the depth of 
the WS well, which is determined to reproduce the experimental separation 
energy, increases. A deeper WS well produces
narrower wave functions, as the curves of the upper panels of
Fig.~\ref{fig:densca} show.

The HF wave functions have a different behavior, as shown in panels (c) and (d). 
In this case, the narrower wave functions are those obtained for the isotopes 
with smaller neutron numbers.  The behavior of the relativistic wave functions 
is somewhat different and does not have a defined trend as in the previous 
cases. It is, in any case, more similar to that of the HF than to that of the 
WS wave functions. 
We point out that the values of the separation energies, and their trend
as a function of the neutron number, are similar in all three types
of calculation.

These differences in the wave functions are responsible for only a
part of the differences in the reduced cross sections of
Figs.~\ref{fig:caiso_par} and \ref{fig:caiso_per}, which show a
different trend as a function of neutron number.  While in the PWIA
the reduced cross section contains only information on the bound-state
wave function, in the DWIA this information is modified by the contribution of the 
other ingredients of the model, such as FSI and the electron-nucleon 
interaction. All these contributions are intertwined in the calculated cross 
section and, in general, they cannot be easily disentangled.

In order to understand the source of the difference between the
behavior of the wave functions and of the cross sections with respect
to the increase of neutron number, we have performed PWIA
calculations. In this case, FSI are switched off and the \eep cross
section is directly proportional to the squared modulus of the Fourier
transform of the hole s.p. wave function. We have checked that the
trend of our PWIA results agrees with the trend of the wave functions
shown in Fig.~\ref{fig:densca}: the WS wave functions produce PWIA
reduced cross sections higher for $^{40}$Ca than for the other nuclei,
while the situation is reversed with the HF wave functions. As a next
step, we have done DWIA calculations by using the same parameters of
the optical potential for all the isotopes. The results are different
from those obtained in the PWIA, but the trends, with respect to the
change of the neutron number, are preserved. Only when we include that
A-dependence of the optical potential are the trends of the cross sections 
changed and also the DWIA-HF results produce $^{40}$Ca \eep 
reduced cross sections larger than those obtained for the other
isotopes.

The dependence of the wave functions on the  proton to neutron 
asymmetry is responsible for a large part of the differences in the reduced cross
sections, but an important and crucial contribution is given by FSI,
which are described in the calculations by phenomenological optical
potentials.  The differences between the results in parallel and
($\q,\omega$) constant kinematics in Figs.~\ref{fig:caiso_par} and
\ref{fig:caiso_per} are basically due to the different effects of the
distortion produced by the optical potential. These effects strongly
depend on kinematics and are larger in parallel than in ($\q,\omega$)
constant kinematics \cite{bof93,bof96,ber82}.

The s.p. bound states adopted in the present calculations are
normalized to unity and no reduction factor has been applied to the
results shown in Figs.~\ref{fig:caiso_par} and \ref{fig:caiso_per} .
The comparison with data in Figs.~\ref{fig:ca40exp} and
\ref{fig:ca48exp} gives, however, a significant quenching of the
measured cross sections with respect to the predictions of the MF
model. The quenching is different for the $^{40}$Ca\eep\ and
$^{48}$Ca\eep\ reactions and increases with neutron number. A
quenching depending on the number of neutrons can be expected for all
the isotopes and would give further differences on the reduced cross
sections than those shown in Figs.~\ref{fig:caiso_par} and
\ref{fig:caiso_per}.

The behavior of the \eep cross sections on nuclei with increasing neutron
number that we have just presented is not specific to the calcium isotopes. An
analogous behavior is found also for  the oxygen isotopes. 
In our calculations we have considered the $^{16,22,24,28}$O nuclei. We show in
Fig.~\ref{fig:oiso} the reduced cross sections of the \eep\ reaction, 
calculated for the emission of a $1p_{1/2}$ proton in these nuclei by using 
our three models. The calculations have been done in the same 
parallel kinematics as in Fig.~\ref{fig:oexp}, and we included the small
differences due to the different proton separation energies of the
different isotopes, as explained in the case of the calcium isotopes.

\begin{figure}
\begin{center}
\includegraphics[height=15cm, width=12cm]{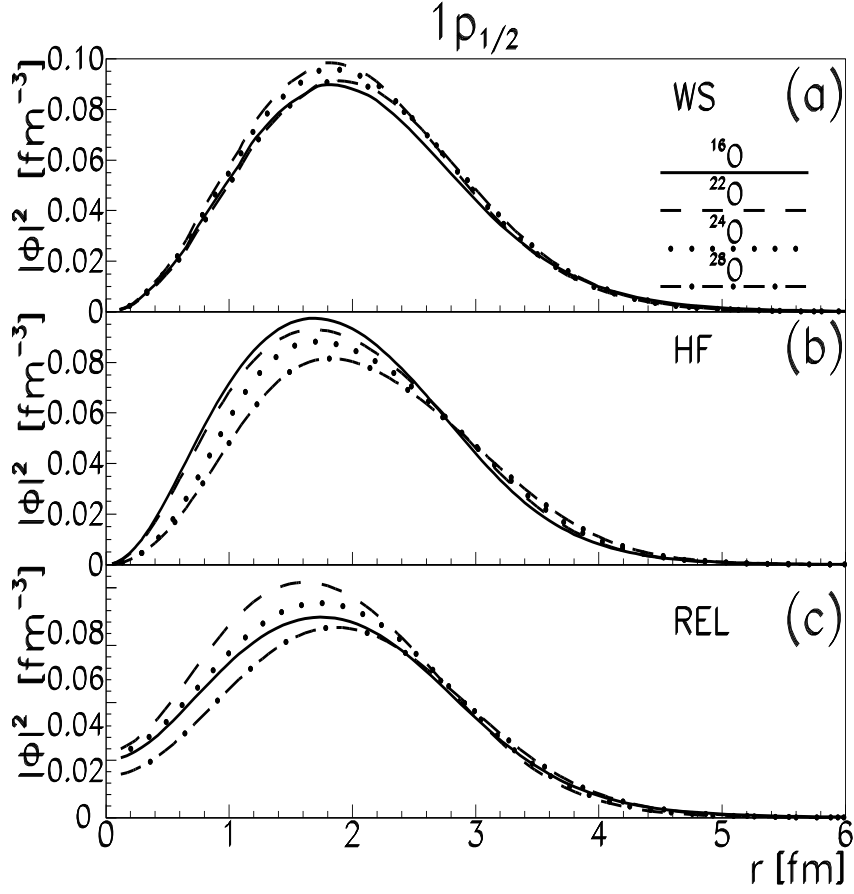} 
\caption{\small 
Squared moduli of the radial part of the $1p_{1/2}$ wave functions for 
$^{16}$O (solid lines), 
$^{22}$O (dashed lines), $^{24}$O (dotted lines), and $^{28}$O (dot-dashed 
lines). In  panel (a)  we show the WS wave functions, 
in panel (b) the HF wave functions, and in
panel (c) the relativistic wave functions obtained in the 
Dirac-Hartree approach.  The normalization of the curves is 
$\int \,dr\, r^2\, |\phi|^2 = 1$. 
}
\label{fig:denso}
\end{center}
\end{figure}

The radial s.p wave functions squared are plotted in
Fig.~\ref{fig:denso}. Also in this case, the different behavior of
the WS and HF radial wave functions with respect to the change in 
neutron number is evident. This difference is not present in the results of 
Fig.~\ref{fig:oexp}. We have repeated also for the oxygen
isotopes the investigation done for the calcium isotopes by doing PWIA
calculations and DWIA calculations with the same optical potential for
all the isotopes. We obtained results analogous to those discussed for
the calcium isotopes. 


\section{Summary and conclusions}
\label{sec:conclusions}

In this work we have presented and discussed \eep cross sections
calculated for a set of calcium and oxygen isotopes with the aim of
studying their evolution with respect to the change of neutron
number.  The calcium and oxygen isotopes have been chosen since data
are available for the doubly magic $^{40}$Ca, $^{48}$Ca, and $^{16}$O
nuclei. In the calcium and oxygen isotope chains, we have considered
only those nuclei with fully occupied s.p. levels, since they are
spherical and, moreover, pairing effects are negligible.

The general framework that we have considered for the description of
the \eep process is based on the one-nucleon knock out picture
\cite{bof82,bof96} and it is well established. The nonrelativistic
DWIA and relativistic RDWIA models used for the calculations were
widely and successfully applied to the analysis of the available \eep
data over a wide range of stable nuclei.  The results obtained with
three different descriptions of the hole wave function of the knocked
out proton, all of them based on the mean field approximation, have
been compared.  We have performed DWIA calculations with
phenomenological WS and HF wave functions. For $^{16}$O, $^{40}$Ca, and
$^{48}$Ca nuclei the WS wave functions are the same as used in
the original analyses of the experimental data in
\cite{kra90t,kra01,leu94}.  The wave functions used in the RDWIA are
obtained by solving Dirac-Hartree equations.

The three models are all able to give a good and similar description
of the shape of the experimental reduced cross section on
$^{16}$O,$^{40}$Ca, and $^{48}$Ca target nuclei, with the only
exception of the HF result in $^{16}$O.  In order to reproduce the
magnitude of the experimental data, a reduction factor has been
applied to all the calculated results.

In our study of the \eep process in the two isotope chains we have
found that the general behavior of the cross sections with respect to
the increasing number of neutrons is analogous for all the three
models.  Generally speaking, the reduced cross sections are larger and
narrower for the lighter isotopes, and evolve by lowering and widening
with increasing neutron number.

The behavior of the hole s.p. wave functions for the three models
show rather different trends: the WS wave functions become narrower
and have higher maxima with increasing neutron number, the HF wave
functions show an opposite behavior, and the behavior of the
relativistic wave functions is not so well defined, but it is more
similar to the behavior of HF than of the WS wave functions.

The dependence of the wave functions on the proton to neutron 
asymmetry is responsible for only a part of the
differences in the reduced cross sections. An important and crucial
contribution is also given by FSI, that are described in the
calculations by phenomenological optical potentials. The optical
potential is an important ingredient of the model that affects both
the size and the shape of the cross section in a way that strongly
depends on kinematics.  In particular, its imaginary part, that  gives a 
reduction of the nucleon flux and, consequently, of the calculated cross 
section, can affect the values of the
spectroscopic factors obtained from the comparison between data and
theoretical results.  In the present calculations we have used
phenomenological optical potentials that were usually adopted in the
previous DWIA and RDWIA analyses of \eep\ data on stable closed-shell
nuclei.  The dependence of the optical potential on the asymmetry
between the number of neutrons and protons is an interesting problem
that deserves careful investigation.

\begin{figure}
\begin{center}
\includegraphics[height=15cm, width=15cm]{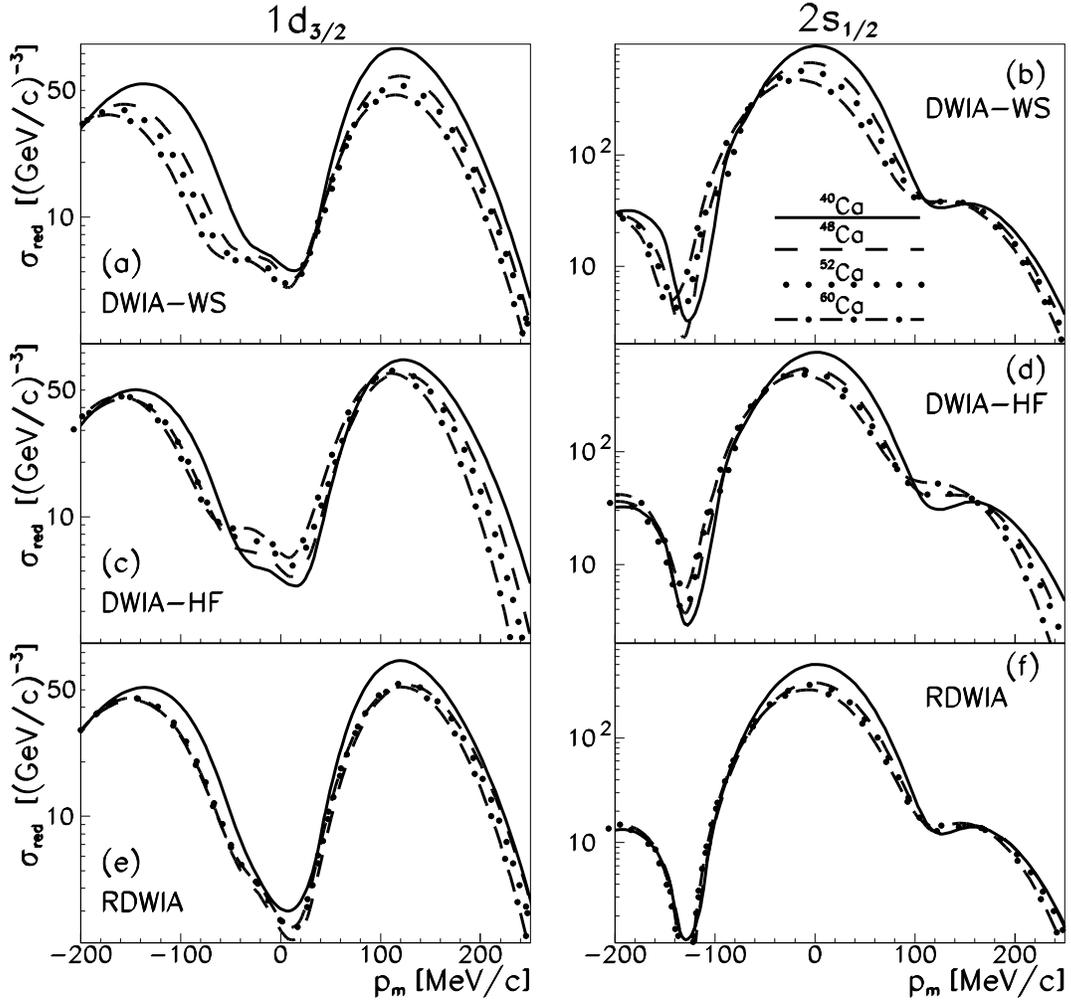} 
\caption{\small Reduced cross section of the \eep\ reaction for
$1d_{3/2}$ (left panels) and $2s_{1/2}$ (right panels) knockout from
$^{40}$Ca (solid lines), $^{48}$Ca (dashed lines), $^{52}$Ca (dotted
lines), and $^{60}$Ca (dot-dashed lines), as a function of $p_{\mathrm
m}$. The results of the DWIA-WS calculations are presented in the  panels (a) 
and (b), and those of the DWIA-HF calculations in the panels (c) and (d). 
The RDWIA results are shown in the panels (e) and (f).
The calculations are done in parallel kinematics with $E_0= 440$ MeV
and $T'= 100$ MeV.
}
\label{fig:caiso_par}
\end{center}
\end{figure}

\begin{figure}
\begin{center}
\includegraphics[height=15cm, width=15cm]{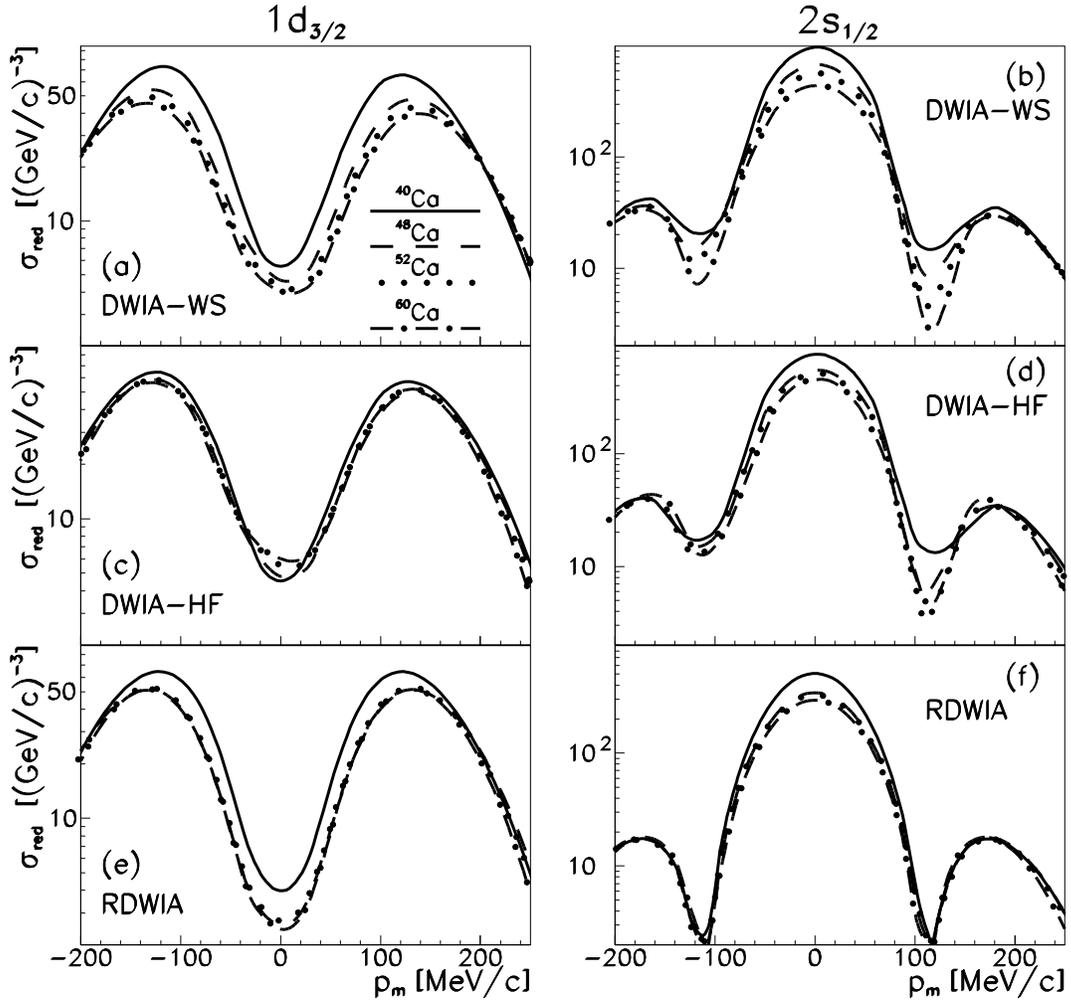} 
\caption{\small The same as in Fig.~\ref{fig:caiso_par} but in
  perpendicular kinematics with $E_0= 483.2$ MeV, $\vartheta=
  61.52^{\circ}$, and $q=450$ MeV/$c$.  }
\label{fig:caiso_per}
\end{center}
\end{figure}

\begin{figure}
\begin{center}
\includegraphics[height=15cm, width=12cm]{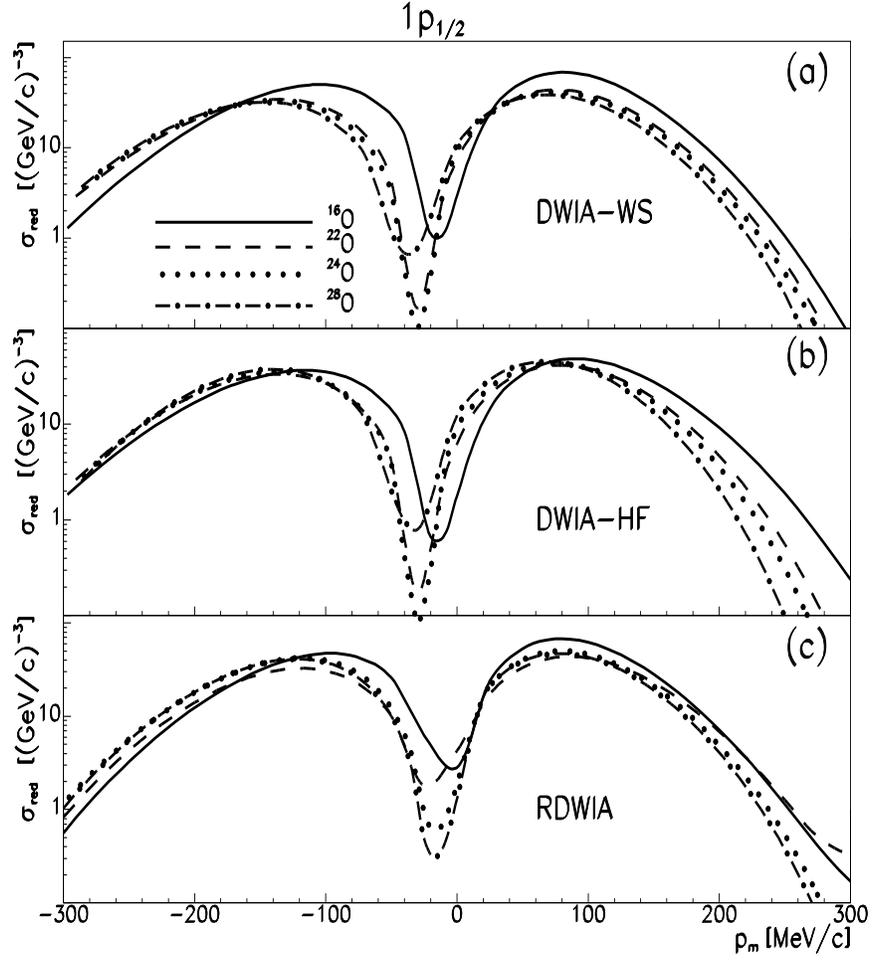} 
\caption{\small Reduced cross section of the \eep\ reaction for  
$1p_{1/2}$ proton 
knockout from  $^{16}$O (solid lines), $^{22}$O (dashed lines), $^{24}$O
(dotted lines), and $^{28}$O (dot-dashed lines) as a function
of $p_{\mathrm m}$. In panel (a) we show the DWIA-WS results, in 
panel (b) the DWIA-HF results, and in panel
(c) the RDWIA results. The calculations have been done in parallel kinematics, 
with $E_0= 520.6$ MeV and $T'= 90$ MeV.  }
\label{fig:oiso}
\end{center}
\end{figure}

We have used MF wave functions
that do not include correlations. The reduction factors found in comparison with 
the experimental data can provide a measure of the effects 
not included in the calculations and, under certain conditions, can be
identified with the spectroscopic factors. In the comparison 
of our \eep cross sections calculated for the various calcium and oxygen 
isotopes, and shown in Figs.~\ref{fig:caiso_par}, \ref{fig:caiso_per}, and 
\ref{fig:oiso}, we did not apply reduction factors, since data are available 
only for the $^{16}$O, $^{40}$Ca, and $^{48}$Ca target nuclei. 

From recent experimental and theoretical studies there are indications
that the spectroscopic factors and the effects of correlations depend
on the asymmetry between the number of neutrons and protons.  Recent
experimental information on the spectroscopic factors of drip-line
isotopes has been obtained by means of nucleon knockout using
intermediate heavy-ion beams \cite{han03,gad04,gad08}.  These results
also include neutron data and suggest a strong dependence of the
spectroscopic factors on the proton-neutron ratio. In general, the
quenching of quasiparticle orbits, and hence correlations, becomes
stronger with increasing separation energy.

From the theoretical point of view, a dependence on nucleon asymmetry
has been found in \cite{rio09} for the depletion of the Fermi sea in
asymmetric nuclear matter, with the minority nucleonic species becoming more 
depleted, and the majority
one less. Results of Faddeev-random=phase approximation calculations for the 
spectroscopic factors of $^{16-28}$O and $^{40-60}$Ca isotopes \cite{bar09a} 
show that the spectroscopic factors
become smaller with increasing nucleon separation energy.  From recent
work on the dispersive optical potential, which has been applied to
calcium isotopes with the aim to extract the asymmetry dependence of
the self-energy, a clear signal emerges suggesting that the surface
imaginary term of the dispersive optical potential increases with the
asymmetry for protons \cite{cha06,cha07,dic10}. Apart from the
associated increased binding that protons experience with increasing
asymmetry, there is a corresponding reduction of valence hole
spectroscopic factors.

Therefore, we can expect a dependence of the spectroscopic factors on
the neutron number and this would give further differences in the
reduced cross sections than those shown in Figs.~\ref{fig:caiso_par},
\ref{fig:caiso_per}, and \ref{fig:oiso}.

Measurements of the exclusive quasifree \eep cross section on nuclei
with neutron excess would offer a unique opportunity for studying the
dependence of the properties of bound protons and of nucleon-nucleon
correlations on the neutron to proton asymmetry. In this work models
that have proven their reliability in the comparison with \eep data on
stable nuclei have been used to investigate the evolution of the \eep
cross sections with increasing proton-neutron asymmetry. Although the
models and the theoretical ingredients adopted in the calculations
contain approximations, our results can serve as a useful first
reference for possible future experiments. The comparison with data
can confirm or invalidate the predictions of our models and test the
ability of the established nuclear theory in the domain of exotic
nuclei.

\newpage
\bibliographystyle{elsart-num} 

\clearpage
\newpage

%
\begin{table}
\begin{ruledtabular}
\begin{tabular}{lcccccc}
& &  $S_p$ [MeV] 
 & r$_{{rms}}$ [fm] & r$_{{rms}}^P$ [fm]& r$_{{rms}}^N$ [fm]
 \\[1mm]
nucleus & &  $1p_{1/2}$
 & $1p_{1/2}$  & & 
 \\[1mm]
 \hline 
  & & &  & & & \\[1mm]
$^{16}$O & WS &    -12.127 & 2.94  & & \\[1mm]
& HF&    -11.906 & 2.83   &2.63 & 2.61 \\[1mm]
& REL&    -12.294 & 2.97  & 2.63& 2.60 \\[1mm]
%
$^{22}$O & WS    & -23.26 & 2.77  & &  \\[1mm]
& HF&    -23.60 &2.83   &2.68 &2.97   \\[1mm]
& REL&   -23.59 & 2.76   &2.59 & 3.01 \\[1mm]
%
$^{24}$O & WS    & -26.60 & 2.77  & &  \\[1mm]
& HF&    -25.60 & 2.89  &2.70 & 3.12 \\[1mm]
& REL&   -23.97 & 2.83   &2.62 &3.28 \\[1mm]
%
$^{28}$O & WS    & -31.04 & 2.80  & & & \\[1mm]
& HF&    -31.04 & 2.96  & 2.82&3.42  \\[1mm]
& REL&   -28.65 & 2.95 &2.69 &3.58   \\[1mm]
\end{tabular}
\caption{\small Proton separation energies ($S_p$), rms radii of the $1p_{1/2}$ bound
state wave functions (r$_{rms}$) and of the global proton (r$_{{rms}}^P$) and 
neutron (r$_{{rms}}^N$) density distributions, obtained with the different MF 
models for all the oxygen isotopes we have considered.  
}
\label{tab:oisot}
\end{ruledtabular}
\end{table}

%
\begin{table}
\begin{ruledtabular}
\begin{tabular}{lccccccc}
 & &  $S_p$ [MeV] 
 & r$_{{rms}}$ [fm]&$S_p$ [MeV] 
 & r$_{{rms}}$ [fm] & r$_{{rms}}^P$ [fm]& r$_{{rms}}^N$ [fm]
 \\[1mm]
nucleus & &  $1d_{3/2}$
 & $1d_{3/2}$ & $2s_{1/2}$   & $2s_{1/2}$ & & 
 \\[1mm]
 \hline 
  & & &  & & & &\\[1mm]
$^{40}$Ca & WS &    -8.328 & 3.69 & -10.85 & 3.72 & \\[1mm]
& HF&    -8.753 & 3.58  &-9.92 & 3.59 &3.37   &3.33   \\[1mm]
& REL&    -8.704 & 3.73  &-9.20 & 3.87 &3.38 & 3.33 \\[1mm]
%
$^{48}$Ca & WS    & -15.907 & 3.54 & -15.807 & 3.58 & & \\[1mm]
& HF&    -16.542 & 3.63 &    -16.18 & 3.55 & 3.41  & 3.55   \\[1mm]
& REL&   -15.606  & 3.60 &   -13.815  & 3.64  &3.37 & 3.59\\[1mm]
%
$^{52}$Ca & WS    & -17.80 & 3.54 & -20.0 & 3.58& & \\[1mm]
& HF&    -18.83 & 3.67 &    -19.98 & 3.55 & 3.46 &  3.70 \\[1mm]
& REL&   -16.75 & 3.68 &   -16.494 & 3.72& 3.41 & 3.85 \\[1mm]
%
$^{60}$Ca & WS    & -25.200 & 3.53  & -24.96 & 3.53 & & \\[1mm]
& HF&    -25.200 & 3.83 &    -24.96 & 3.64  &3.59  &  3.98 \\[1mm]
& REL&   -20.744  & 3.75  &   -20.288  & 3.76 &3.51 & 4.15 \\[1mm]
\end{tabular}
\caption{\small 
The same as in Tab. \ref{tab:oisot} for the
$1d_{3/2}$ and $2s_{1/2}$ proton states for all the calcium isotopes
we have considered.
}
\label{tab:caisot}

\end{ruledtabular}
\end{table}

%
\begin{table}
\begin{ruledtabular}
\begin{tabular}{lccccc}
                       &    & DWIA-WS & DWIA-HF  & RDWIA   &  \\ \hline
($3/2^{~+}~;~^{39}$K)  & $\parallel$ & 0.65 & 0.64 & 0.69 &
                       Fig. \ref{fig:ca40exp} \\
($3/2^{~+}~;~^{39}$K)  & $\perp$ & 0.49 & 0.51 & 0.49 & Fig. \ref{fig:ca40exp} \\
($1/2^{~+}~;~^{39}$K)  & $\parallel$ & 0.52 & 0.57 & 0.51 &
                       Fig. \ref{fig:ca40exp} \\
($1/2^{~+}~;~^{39}$K)  & $\perp$ & 0.55 & 0.62 & 0.51 & Fig. \ref{fig:ca40exp} \\
($3/2^{~+}~;~^{47}$K)  & $\parallel$ & 0.56 & 0.55 & 0.52 &
                       Fig. \ref{fig:ca48exp} \\
($1/2^{~+}~;~^{47}$K)  & $\parallel$ & 0.54 & 0.58 & 0.55 &
                       Fig. \ref{fig:ca48exp} \\
($1/2^{~-}~;~^{15}$N)  & $\parallel$ & 0.64 & 0.89 & 0.70 & Fig. \ref{fig:oexp} \\
\end{tabular}
\caption{\small Reduction factors applied to the calculated reduced
  cross sections. We have
  indicated the spin and parity of the state of the residual nucleus.
  The symbols $\parallel$ and $\perp$ indicate the parallel and
  perpendicular or ($\q,\omega$) constant kinematics, respectively.  
}
\label{tab:redfac}
\end{ruledtabular}
\end{table}
\clearpage
\newpage
\end{document}